\documentclass[12pt,a4paper]{article}
\usepackage{cite}
\usepackage{latexsym}
\usepackage{amsbsy}
\usepackage{amssymb}
\usepackage[dvips]{graphicx}
\advance \textheight by \topskip
\begin{document}

\begin{center}
\Large \bfseries
Four-neutrino mixing
\\[0.3cm]
\large \mdseries \upshape
C. Giunti
\\
\normalsize \itshape
INFN, Sezione di Torino, 
and
Dipartimento di Fisica Teorica,
\\
Universit\`a di Torino,
Via P. Giuria 1, I--10125 Torino, Italy, and
\\
School of Physics, Korea Institute for Advanced Study,
Seoul 130-012, Korea
\end{center}
\begin{abstract}
It is shown that only two four-neutrino
schemes are compatible with the results
of all neutrino oscillation experiments.
These two schemes have a mass spectrum composed of
two pairs of neutrinos with close masses 
separated by the ``LSND gap'' of the order of 1 eV.
Talk presented at the
X$^{\mathrm{th}}$ International School ``PARTICLES and COSMOLOGY'',
19--25 April 1999,
Baksan Valley, Kabardino-Balkaria, Russia,
KIAS-P99067,
hep-ph/9907485.
\end{abstract}

The investigation of neutrino properties is considered today
as one of the most exciting fields of research in high-energy physics.
It has received a stupendous boost from the evidence in favor of neutrino
oscillations discovered recently in the
Super-Kamiokande experiment \cite{SK-atm,Scholberg-99}.

The main Super-Kamiokande evidence in favor of neutrino
oscillations consists in
the observation of an up--down asymmetry of high-energy
$\mu$-like events generated by atmospheric neutrinos:
\begin{equation}
\mathcal{A}_\mu
\equiv
(D_\mu-U_\mu)/(D_\mu+U_\mu)
=
0.311 \pm 0.043 \pm 0.01
\qquad
\mbox{\protect\cite{Scholberg-99}}
\,.
\label{Amu}
\end{equation}
Here $D_\mu$ and $U_\mu$ are,
respectively,
the number of downward-going and upward-going events,
corresponding to the zenith angle intervals
$0.2 < \cos\theta < 1$
and
$-1 < \cos\theta < -0.2$.
Since the fluxes of high-energy downward-going and upward-going
atmospheric neutrinos are predicted to be equal with high accuracy
on the basis of geometrical arguments
(see \cite{Lipari-99}),
the Super-Kamiokande evidence in favor of neutrino
oscillations is model-independent and
provides a definite confirmation of the indications
in favor of oscillations of atmospheric neutrinos
found in the Super-Kamiokande experiment itself
\cite{SK-atm,Scholberg-99,SK-atm-upmu}
and in other experiments through the measurement
of the ratio of $\mu$-like and $e$-like events
(Kamiokande,
IMB,
Soudan 2 \cite{atm-exp-contained})
and through the measurement
of upward-going muons produced by neutrino interactions
in the rock below the detector
(MACRO \cite{MACRO}).
Large
$\nu_\mu\leftrightarrows\nu_e$
oscillations of atmospheric neutrinos
are excluded
by the absence of a
up--down asymmetry of high-energy
$e$-like events generated by atmospheric neutrinos and detected in
the Super-Kamiokande experiment
($
\mathcal{A}_e
=
0.036 \pm 0.067 \pm 0.02
$
\cite{SK-atm})
and by the negative result
of the CHOOZ long-baseline $\bar\nu_e$ disappearance experiment
\cite{CHOOZ-99}.
Therefore,
the atmospheric neutrino anomaly consists in the disappearance
of muon neutrinos and can be explained by
$\nu_\mu\to\nu_\tau$
and/or
$\nu_\mu\to\nu_s$
oscillations
(here $\nu_s$ is a sterile neutrino
that does not take part in weak interactions).

Other indications in favor of neutrino oscillations have been obtained in
solar neutrino experiments
(Homestake,
Kamiokande,
GALLEX,
SAGE,
Super-Kamiokande \cite{sun-exp})
and in the LSND experiment \cite{LSND}.

The flux of electron neutrinos measured in
all five solar neutrino experiments
is substantially smaller than the one predicted
by the Standard Solar Model \cite{BP98}
and a comparison of the data of different experiments
indicate an energy dependence of the solar $\nu_e$ suppression,
which represents a rather convincing evidence
in favor of neutrino oscillations
(see \cite{BGG-98-review} and references therein).
The disappearance of solar
electron neutrinos
can be explained by
$\nu_e\to\nu_\mu$
and/or
$\nu_e\to\nu_\tau$
and/or
$\nu_e\to\nu_s$
oscillations
(see the recent analyses in \cite{sun-analysis}).

The accelerator LSND experiment
is the only one that claims the observation
of neutrino oscillations in specific appearance channels:
$\bar\nu_\mu\to\bar\nu_e$ and $\nu_\mu\to\nu_e$.
Since
the appearance of neutrinos with a different flavor
represents
the true essence of neutrino oscillations,
the LSND evidence is extremely interesting
and its confirmation (or disproof)
by other experiments
should receive high priority in future research.
Four such experiments have been proposed and are under study:
BooNE at Fermilab,
I-216 at CERN,
ORLaND at Oak Ridge
and
NESS at the European Spallation Source \cite{LSND-check}.
Among these proposals only BooNE is approved and will start in 2001.

Neutrino oscillations occur
if neutrinos are massive and mixed particles
(see \cite{reviews,BGG-98-review}),
\textit{i.e.} if
the left-handed components
$\nu_{{\alpha}L}$
of the flavor neutrino fields
are superpositions of
the left-handed components
$\nu_{kL}$
($k=1,\ldots,N$)
of neutrino fields with definite mass
$m_k$:
\begin{equation}
\nu_{{\alpha}L}
=
\sum_{k=1}^{N}
U_{{\alpha}k}
\,
\nu_{kL}
\,,
\label{mixing}
\end{equation}
where $U$
is a $N{\times}N$ unitary mixing matrix.
From the measurement of the invisible decay width of the $Z$-boson
(see \cite{PDG98})
it is known hat the number of light active
neutrino flavors is three,
corresponding to $\nu_e$, $\nu_\mu$ and $\nu_\tau$
(active neutrinos are
those taking part to standard weak interactions).
This implies that
the number $N$ of massive neutrinos is bigger or equal to three.
If $N>3$, in the flavor basis there are $N_s=N-3$
sterile neutrinos,
$\nu_{s_1}$, \ldots, $\nu_{s_{N_s}}$,
that do not take part to standard weak interactions.
In this case the index $\alpha$ in Eq. (\ref{mixing})
takes the values
$e,\mu,\tau,s_1,\ldots,s_{N_s}$.

The three evidences in favor of neutrino oscillations
found in solar and atmospheric neutrino experiments
and in the accelerator LSND experiment
imply the existence of at least three independent
neutrino mass-squared differences.
This can be seen by considering
the general expression for the probability of
$\nu_\alpha\to\nu_\beta$
transitions in vacuum,
that can be written as
(see \cite{reviews})
\begin{equation}
P_{\nu_\alpha\to\nu_\beta}
=
\left|
\sum_{k=1}^{N}
U_{{\alpha}k}^* \,
U_{{\beta}k} \,
\exp\left( - i \, \frac{ \Delta{m}^2_{kj} \, L }{ 2 \, E } \right)
\right|^2
\,,
\label{Posc}
\end{equation}
where
$ \Delta{m}^2_{kj} \equiv m_k^2-m_j^2 $,
$j$ is any of the mass-eigenstate indices,
$L$ is the distance between the neutrino source and detector
and $E$ is the neutrino energy.
The range of $L/E$ characteristic of each type of experiment is different:
$ L / E \gtrsim 10^{10} \, \mathrm{eV}^{-2} $
for solar neutrino experiments,
$ L / E \sim 10^{2} - 10^{3} \, \mathrm{eV}^{-2} $
for atmospheric neutrino experiments
and
$ L / E \sim 1 \, \mathrm{eV}^{-2} $
for the LSND experiment.
From Eq. (\ref{Posc}) it is clear that neutrino oscillations
are observable in an experiment only if there is at least one
mass-squared difference
$\Delta{m}^2_{kj}$ such that
\begin{equation}
\frac{ \Delta{m}^2_{kj} \, L }{ 2 \, E }
\gtrsim 0.1
\label{cond1}
\end{equation}
(the precise lower bound depends on the sensitivity of the experiment)
in a significant part of the energy and source-detector distance
intervals of the experiment
(if the condition (\ref{cond1}) is not satisfied,
$
P_{\nu_\alpha\to\nu_\beta}
\simeq
\left|
\sum_k
U_{{\alpha}k}^* \,
U_{{\beta}k}
\right|^2
=
\delta_{\alpha\beta}
$).
Since the range of $L/E$ probed by the LSND experiment is the smaller one,
a large mass-squared difference is needed for LSND oscillations:
\begin{equation}
\Delta{m}^2_{\mathrm{LSND}} \gtrsim 10^{-1} \, \mathrm{eV}^2
\,.
\label{dm2-LSND}
\end{equation}
Furthermore,
from Eq. (\ref{Posc}) it is clear that 
a dependence of the oscillation probability
from the neutrino energy $E$
and the source-detector distance $L$
is observable only if there is at least one
mass-squared difference
$\Delta{m}^2_{kj}$ such that
\begin{equation}
\frac{ \Delta{m}^2_{kj} \, L }{ 2 \, E }
\sim 1
\,.
\label{cond2}
\end{equation}
Indeed,
all the phases
$ \Delta{m}^2_{kj} L / 2 E \gg 1 $
are washed out by the average over the energy and source-detector
ranges characteristic of the experiment.
Since
a variation of the oscillation probability as a function of neutrino energy
has been observed both in solar and atmospheric neutrino experiments
and the ranges of $L/E$ characteristic
of these two types of experiments are different from each other
and different from the LSND range,
two more mass-squared differences with different scales are needed:
\begin{eqnarray}
&&
\Delta{m}^2_{\mathrm{sun}} \sim 10^{-10} \, \mathrm{eV}^2
\qquad
\mbox{(VO)}
\,,
\label{dm2-sun}
\\
&&
\Delta{m}^2_{\mathrm{atm}} \sim 10^{-3} - 10^{-2} \, \mathrm{eV}^2
\,.
\label{dm2-atm}
\end{eqnarray}
The condition (\ref{dm2-sun}) for the solar mass-squared difference
$\Delta{m}^2_{\mathrm{sun}}$
has been obtained under the assumption of vacuum oscillations (VO). 
If the disappearance of solar $\nu_e$'s is due to the MSW effect \cite{MSW},
the condition
\begin{equation}
\Delta{m}^2_{\mathrm{sun}} \lesssim 10^{-4} \, \mathrm{eV}^2
\qquad
\mbox{(MSW)}
\label{dm2-sun-MSW}
\end{equation}
must be fulfilled
in order to have a resonance in the interior of the sun.
Hence,
in the MSW case
$\Delta{m}^2_{\mathrm{sun}}$
must be at least one order of magnitude smaller than
$\Delta{m}^2_{\mathrm{atm}}$.

It is possible to ask if three different scales
of neutrino mass-squared differences are needed even
if the results of the Homestake solar neutrino experiment
is neglected, allowing an energy-independent suppression of the solar $\nu_e$ flux.
The answer is that still the data cannot be fitted with only
two neutrino mass-squared differences
because an energy-independent suppression of the solar $\nu_e$ flux
requires large $\nu_e\to\nu_\mu$ or $\nu_e\to\nu_\tau$
transitions generated by $\Delta{m}^2_{\mathrm{atm}}$
or $\Delta{m}^2_{\mathrm{LSND}}$.
These transitions
are forbidden by the results of the Bugey \cite{Bugey-95}
and CHOOZ \cite{CHOOZ-99}
reactor $\bar\nu_e$ disappearance experiments
and by the non-observation of an up-down asymmetry
of $e$-like events in the Super-Kamiokande
atmospheric neutrino experiment \cite{SK-atm}.

The existence of three different scales of $\Delta{m}^2$
imply that at least four light massive neutrinos must exist in nature.
Here we consider the schemes with four light and mixed neutrinos
\cite{four-models,four-phenomenology,BGKP-96,%
BGG-AB-96,Okada-Yasuda-97,BGG-bounds,BGG-CP,%
BGGS-BBN-98,Barger-variations-98,BGGS-AB-99},
which constitute the minimal possibility
that allows to explain all the existing data with neutrino oscillations.
In this case,
in the flavor basis the three active neutrinos $\nu_e$, $\nu_\mu$, $\nu_\tau$
are accompanied by a sterile neutrino $\nu_s$
that does not take part in
standard weak interactions.

\begin{figure}[t]
\begin{center}
\includegraphics[bb=13 668 522 827,width=0.99\linewidth]{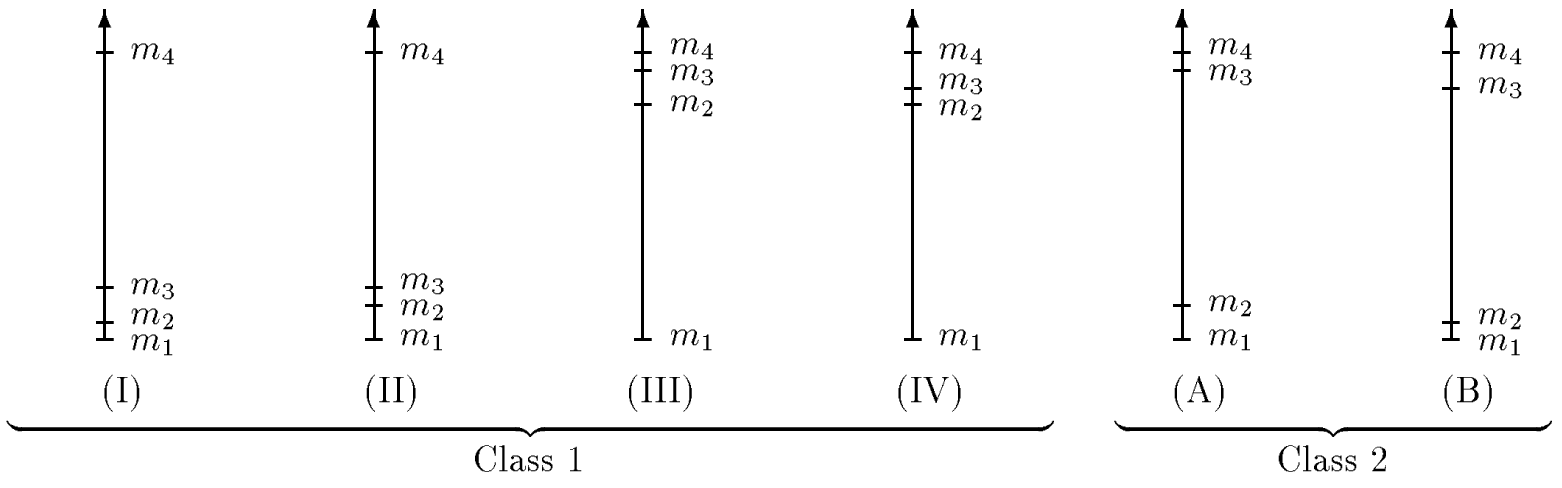}
\refstepcounter{figure}
\label{4spectra}
\small
Figure \ref{4spectra}
\end{center}
\end{figure}

The six types of four-neutrino mass spectra
with
three different scales of $\Delta{m}^2$
that can accommodate
the hierarchy
$
\Delta{m}^2_{\mathrm{sun}}
\ll
\Delta{m}^2_{\mathrm{atm}}
\ll
\Delta{m}^2_{\mathrm{LSND}}
$
are shown in Fig.~\ref{4spectra}.
In all these mass spectra there are two groups
of close masses separated by the ``LSND gap'' of the order of 1 eV.
In each scheme the smallest mass-squared
difference corresponds to
$\Delta{m}^2_{\mathrm{sun}}$
($\Delta{m}^2_{21}$ in schemes I and B,
$\Delta{m}^2_{32}$ in schemes II and IV,
$\Delta{m}^2_{43}$ in schemes III and A),
the intermediate one to
$\Delta{m}^2_{\mathrm{atm}}$
($\Delta{m}^2_{31}$ in schemes I and II,
$\Delta{m}^2_{42}$ in schemes III and IV,
$\Delta{m}^2_{21}$ in scheme A,
$\Delta{m}^2_{43}$ in scheme B)
and the largest mass squared difference
$ \Delta{m}^2_{41} = \Delta{m}^2_{\mathrm{LSND}} $
is relevant for the oscillations observed in the LSND experiment.
The six schemes are divided into four schemes of class 1 (I--IV)
in which there is a group of three masses separated from an isolated mass
by the LSND gap,
and two schemes of class 2 (A, B)
in which there are two couples of close masses separated by the LSND gap.

In the following we will show that
the schemes of class 1 (I--IV) are disfavored by the data
if also the negative
results of short-baseline accelerator and reactor disappearance
neutrino oscillation experiments are taken into account
\cite{BGG-AB-96,Barger-variations-98,BGGS-AB-99}.
Let us remark that in principle one could check which schemes are allowed
by doing a combined fit of all data in the framework of
the most general four-neutrino mixing scheme,
with three mass-squared differences,
six mixing angles and three CP-violating phases
as free parameters.
However,
at the moment it is not possible to perform such a fit
because of the enormous complications due
to the presence of too many parameters
and to the difficulties involved in a combined fit of
the data of different experiments,
which are usually analyzed by the experimental collaborations
using different methods.
Hence,
we think that it is quite remarkable
that one can exclude the schemes of class 1
with the following relatively simple procedure.

Let us define the quantities $d_\alpha$,
with $\alpha=e,\mu,\tau,s$,
in the schemes of class 1 as
\begin{equation}
d_\alpha^{\mathrm{(I)}}
\equiv
|U_{{\alpha}4}|^2
\,,
\qquad
d_\alpha^{\mathrm{(II)}}
\equiv
|U_{{\alpha}4}|^2
\,,
\qquad
d_\alpha^{\mathrm{(III)}}
\equiv
|U_{{\alpha}1}|^2
\,,
\qquad
d_\alpha^{\mathrm{(IV)}}
\equiv
|U_{{\alpha}1}|^2
\,.
\label{dalpha}
\end{equation}
Physically $d_\alpha$ quantifies the mixing of the flavor neutrino $\nu_\alpha$
with the isolated neutrino,
whose mass is separated from the other three by the LSND gap.

The probability of
$\nu_\alpha\to\nu_\beta$
($\beta\neq\alpha$)
and
$\nu_\alpha\to\nu_\alpha$
transitions
(and the corresponding probabilities for antineutrinos)
in short-baseline experiments
are given by \cite{BGG-AB-96,BGG-98-review}
\begin{equation}
P_{\nu_\alpha\to\nu_\beta}
=
A_{\alpha;\beta} \, \sin^2 \frac{ \Delta{m}^2_{41} L }{ 4 E }
\,,
\qquad
P_{\nu_\alpha\to\nu_\alpha}
=
1
-
B_{\alpha;\alpha} \, \sin^2 \frac{ \Delta{m}^2_{41} L }{ 4 E }
\,,
\label{prob}
\end{equation}
with
the oscillation amplitudes
\begin{equation}
A_{\alpha;\beta}
=
4 \, d_\alpha \, d_\beta
\,,
\qquad
B_{\alpha;\alpha}
=
4 \, d_\alpha \, ( 1 - d_\alpha )
\,.
\label{ampli}
\end{equation}
The probabilities (\ref{prob})
have the same form as the corresponding probabilities in the
case of two-neutrino mixing,
$
P_{\nu_\alpha\to\nu_\beta}
=
\sin^2(2\vartheta) \, \sin^2(\Delta{m}^2L/4E)
$
and
$
P_{\nu_\alpha\to\nu_\alpha}
=
1
-
\sin^2(2\vartheta) \, \sin^2(\Delta{m}^2L/4E)
$,
which have been used by all experimental collaborations
for the analysis of the data in order to get information
on the parameters
$\sin^2(2\vartheta)$
and
$\Delta{m}^2$
($\vartheta$ and $\Delta{m}^2$ are, respectively, the mixing angle
and the mass-squared difference in the case of two-neutrino mixing).
Therefore,
we can use the results of their analyses in order to get information
on the corresponding parameters
$A_{\alpha;\beta}$,
$B_{\alpha;\alpha}$
and
$\Delta{m}^2_{41}$.

The exclusion plots obtained in short-baseline
$\bar\nu_e$ and $\nu_\mu$
disappearance experiments
imply that \cite{BGG-AB-96}
\begin{equation}
d_\alpha \leq a^0_\alpha
\qquad \mbox{or} \qquad
d_\alpha \geq 1-a^0_\alpha
\qquad
(\alpha=e,\mu)
\,,
\label{small-large}
\end{equation}
with
\begin{equation}
a_\alpha^0
=
\frac{1}{2}
\left( 1 - \sqrt{ 1 - B_{\alpha;\alpha}^0 } \, \right)
\qquad
(\alpha=e,\mu)
\,,
\label{aa0}
\end{equation}
where
$B_{e;e}^0$
and
$B_{\mu;\mu}^0$
are the upper bounds,
that depend on
$\Delta{m}^2_{41}$,
of the oscillation amplitudes
$B_{e;e}$
and
$B_{\mu;\mu}$
given by the exclusion plots of $\bar\nu_e$ and $\nu_\mu$
disappearance experiments.
From the exclusion curves of the Bugey reactor $\bar\nu_e$
disappearance experiment \cite{Bugey-95}
and of the CDHS and CCFR accelerator $\nu_\mu$ disappearance experiments
\cite{CDHS-CCFR-84}
it follows that
$ a_e^0 \lesssim 4 \times 10^{-2} $
for
$\Delta m^2_{41} \gtrsim 0.1 \, \mathrm{eV}^2$
and
$ a_\mu^0 \lesssim 0.2 $
for
$\Delta m^2_{41} \gtrsim 0.4 \, \mathrm{eV}^2$
(see \cite{BGG-98-review}).

\begin{figure}[t!]
\begin{tabular*}{\linewidth}{@{\extracolsep{\fill}}cc}
\begin{minipage}{0.47\linewidth}
\begin{center}
\includegraphics[bb=60 148 483 562,width=0.99\linewidth]{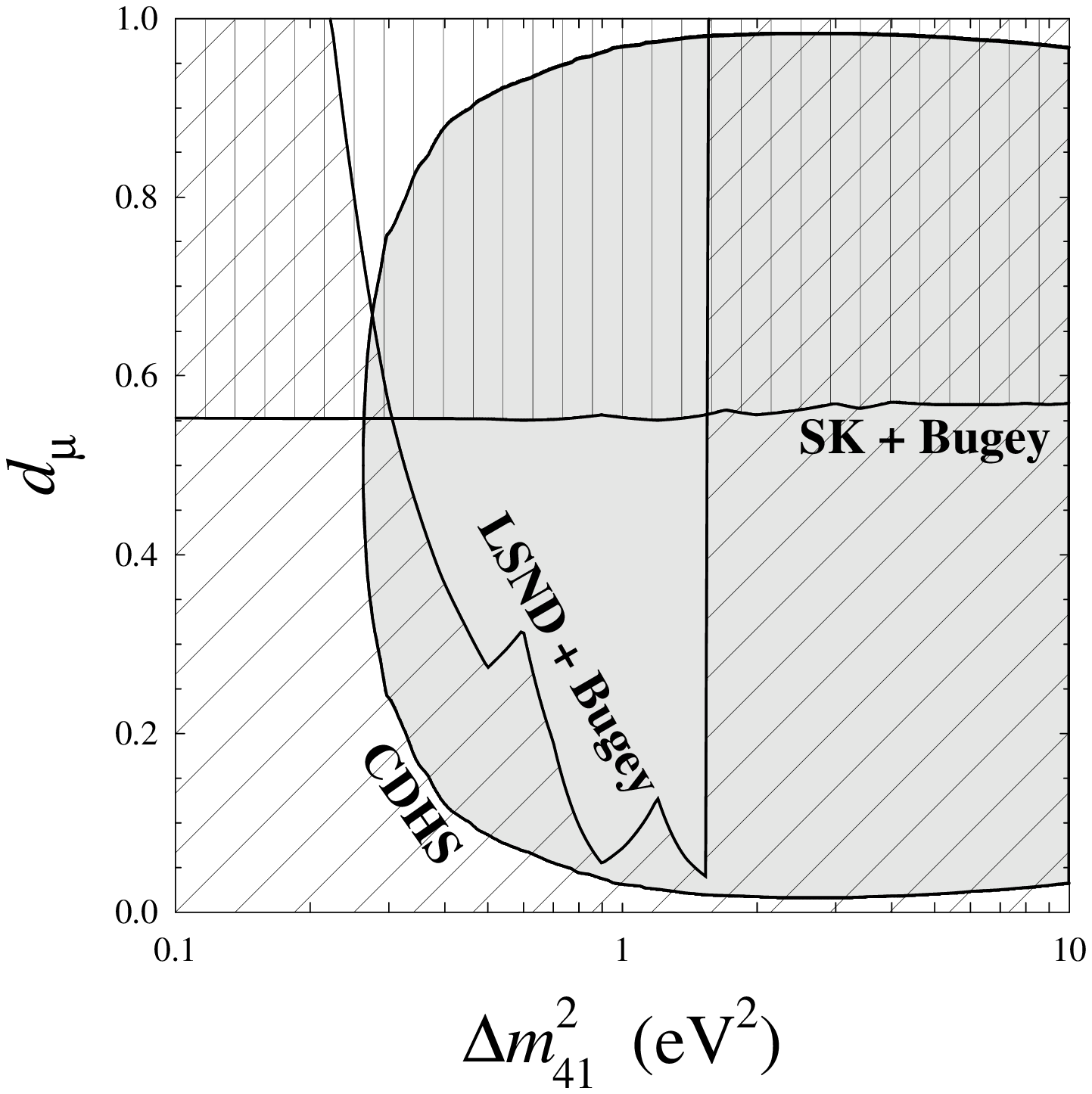}
\refstepcounter{figure}
\label{dmu}
\small
Figure \ref{dmu}
\end{center}
\end{minipage}
&
\begin{minipage}{0.47\linewidth}
\begin{center}
\includegraphics[bb=65 153 485 563,width=0.99\linewidth]{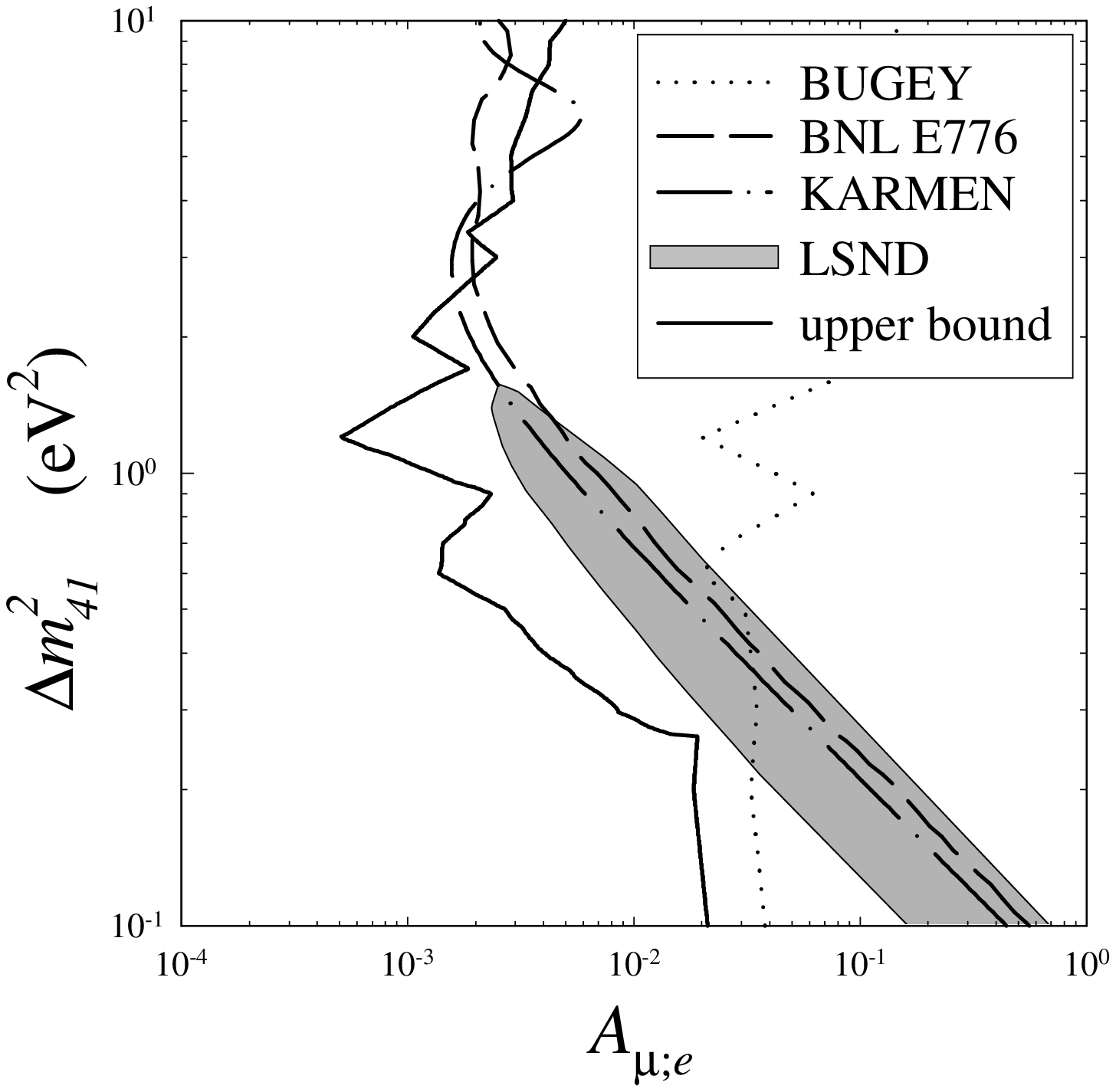}
\refstepcounter{figure}
\label{amuel}
\small
Figure \ref{amuel}
\end{center}
\end{minipage}
\end{tabular*}
\end{figure}

Therefore,
the negative results of
short-baseline
$\bar\nu_e$ and $\nu_\mu$
disappearance experiments
imply that $d_e$ and $d_\mu$
are either small or large (close to one).
However,
since the survival probability of solar $\nu_e$'s is bounded by
\cite{BGG-AB-96,BGG-98-review}
\begin{equation}
P^{\mathrm{sun}}_{\nu_e\to\nu_e} \geq d_e^2
\,,
\label{solar-bound}
\end{equation}
only the possibility
\begin{equation}
d_e \leq a^0_e
\label{de-bound}
\end{equation}
is acceptable in order to explain the observed deficit of solar $\nu_e$'s
with neutrino oscillations.
In a similar way,
since the survival probability of atmospheric $\nu_\mu$'s and $\bar\nu_\mu$'s
is bounded by
\cite{BGG-AB-96,BGG-98-review} 
\begin{equation}
P^{\mathrm{atm}}_{\nu_\mu\to\nu_\mu} \geq d_\mu^2
\,,
\label{atm-bound}
\end{equation}
it is clear that large values of
$d_\mu$
are incompatible with the asymmetry (\ref{Amu})
observed in the Super-Kamiokande experiment.
Indeed, it has been shown in \cite{BGGS-AB-99}
that the Super-Kamiokande asymmetry (\ref{Amu})
and
the exclusion curve of the Bugey $\bar\nu_e$
disappearance experiment
imply the upper bound
\begin{equation}
d_\mu \lesssim 0.55 \equiv a^{\mathrm{SK}}_\mu
\,.
\label{dmu-bound}
\end{equation}
This upper bound is
depicted by the horizontal line in Fig.~\ref{dmu}
(the vertically hatched area above the line is excluded).

In Fig.~\ref{dmu} we have also shown the bound
$d_\mu \leq a^0_\mu$
or
$d_\mu \geq 1-a^0_\mu$
obtained from the exclusion plot of the short-baseline
CDHS
$\nu_\mu$ disappearance experiment,
which exclude the shadowed region.
It is clear that the results of short-baseline disappearance experiments
and the Super-Kamiokande asymmetry (\ref{Amu})
imply that
$ d_\mu \lesssim 0.55 $
for
$ \Delta{m}^2_{41} \lesssim 0.3 \, \mathrm{eV}^2 $
and that
$ d_\mu $
is very small for
$ \Delta{m}^2_{41} \gtrsim 0.3 \, \mathrm{eV}^2 $.
However, this range of $d_\mu$ is disfavored by the results of the LSND
experiment,
that
imply a lower bound
$A^\mathrm{min}_{\mu;e}$
for the amplitude
$A_{\mu;e} = 4 d_e d_\mu$
of $\nu_\mu\to\nu_e$ oscillations.
Using also the bound (\ref{de-bound}),
we obtain the constraint
\begin{equation}\label{LSND}
d_\mu \geq A^\mathrm{min}_{\mu;e}/4a^0_e
\,.
\end{equation}
This bound is represented by the curve in Fig.~\ref{dmu} labelled
LSND + Bugey
(the diagonally hatched area is excluded)
and one can see that it excludes the range of $d_\mu$
allowed by the results of short-baseline disappearance experiments
and by the Super-Kamiokande asymmetry (\ref{Amu}).

From Fig.~\ref{dmu}
one can see that
in the framework of the schemes of class 1
there is no range of $d_\mu$
that is compatible with all the experimental data.
Hence,
the four-neutrino schemes of class 1 are
disfavored by the data.

The incompatibility of the experimental results with the schemes of class 1
is shown also in Fig.~\ref{amuel},
where we have plotted in the $A_{\mu;e}$--$\Delta{m}^2_{41}$ plane
the upper bound
$ A_{\mu;e} \leq 4 \, a^0_e \, a^0_\mu $
for
$ \Delta{m}^2_{41} > 0.26 \, \mathrm{eV}^2 $
and
$ A_{\mu;e} \leq 4 \, a^0_e \, a^{\mathrm{SK}}_\mu $
for
$ \Delta{m}^2_{41} < 0.26 \, \mathrm{eV}^2 $
(solid line, the region on the right is excluded).
One can see that this constraint is incompatible with
the LSND-allowed region
(shadowed area).

On the other hand,
the four-neutrino schemes of class 2 (A, B)
are compatible with the results of all neutrino oscillation experiments
if
the mixing of $\nu_e$ with the two mass eigenstates responsible
for the oscillations of solar neutrinos
($\nu_3$ and $\nu_4$ in scheme A
and
$\nu_1$ and $\nu_2$ in scheme B)
is large
and
the mixing of $\nu_\mu$ with the two mass eigenstates responsible
for the oscillations of atmospheric neutrinos
($\nu_1$ and $\nu_2$ in scheme A
and
$\nu_3$ and $\nu_4$ in scheme B)
is large
\cite{BGKP-96,BGG-AB-96,Barger-variations-98,BGGS-AB-99}.
This fact implies that $\nu_e$'s do not oscillate
in atmospheric and long-baseline neutrino oscillation experiments
and one can obtain rather stringent upper bounds
for the probability of $\nu_e$ transitions into any other state
\cite{BGG-bounds}
and for the size of CP or T violation that could be measured
in long-baseline experiments
in the $\nu_\mu\leftrightarrows\nu_e$
and $\bar\nu_\mu\leftrightarrows\bar\nu_e$ channels
\cite{BGG-CP}.
Furthermore,
it has been shown in \cite{Okada-Yasuda-97,BGGS-BBN-98}
that the upper bound $ N_\nu^{\mathrm{BBN}} < 4 $
for the effective number of neutrinos
in Big-Bang Nucleosynthesis
implies that
the mixing of $\nu_s$
with the two mass eigenstates responsible
for the oscillations of atmospheric neutrinos
is very small.
In this case atmospheric neutrinos oscillate only in the
$\nu_\mu\to\nu_\tau$ channel
and solar neutrino oscillate only in the
$\nu_e\to\nu_s$ channel.
This is very important because it implies that the two-generation
analyses of solar and atmospheric neutrino data
give correct information on neutrino mixing
in the two four-neutrino schemes A and B.

\bigskip

I would like to thank S.M. Bilenky, W. Grimus and C.W. Kim for friendship
and stimulating collaboration.
I would also like to express my gratitude to
the Korea Institute for Advanced Study (KIAS)
for kind hospitality during the writing of this report.


\small
\begin{thebibliography}{10}

\bibitem{SK-atm}
Y. Fukuda \textit{et al.} (Super-Kamiokande Coll.),
Phys. Rev. Lett. \textbf{81}, 1562 (1998).

\bibitem{Scholberg-99}
K. Scholberg (Super-Kamiokande Coll.),
hep-ex/9905016.

\bibitem{Lipari-99}
P. Lipari,
hep-ph/9904443.

\bibitem{SK-atm-upmu}
Y. Fukuda \textit{et al.} (Super-Kamiokande Coll.),
Phys. Rev. Lett. \textbf{82}, 2644 (1999);
A. Habig (Super-Kamiokande Coll.), hep-ex/9903047.

\bibitem{atm-exp-contained}
Y. Fukuda \textit{et al.} (Kamiokande Coll.),
Phys. Lett. \textbf{B335}, 237 (1994);
R. Becker-Szendy \textit{et al.} (IMB Coll.),
Nucl. Phys. B (Proc. Suppl.) \textbf{38}, 331 (1995);
W.W.M. Allison \textit{et al.} (Soudan Coll.),
Phys. Lett. \textbf{B449}, 137 (1999).

\bibitem{MACRO}
M. Ambrosio \textit{et al.} (MACRO Coll.),
Phys. Lett. \textbf{B434}, 451 (1998);
P. Bernardini (MACRO Coll.),
hep-ex/9906019.

\bibitem{CHOOZ-99}
M. Apollonio \textit{et al.} (CHOOZ Coll.),
Phys. Lett. \textbf{B420}, 397 (1998);
hep-ex/9907037.

\bibitem{sun-exp}
B.T. Cleveland \textit{et al.},
Astrophys. J. \textbf{496}, 505 (1998);
K.S. Hirata \textit{et al.} (Kamiokande Coll.),
Phys. Rev. Lett. \textbf{77}, 1683 (1996);
W. Hampel \textit{et al.} (GALLEX Coll.),
Phys. Lett. \textbf{B447}, 127 (1999);
J.N. Abdurashitov \textit{et al.} (SAGE Coll.),
astro-ph/9907113;
Y. Fukuda \textit{et al.} (Super-Kamiokande Coll.),
Phys. Rev. Lett. \textbf{81}, 1158 (1998);
Phys. Rev. Lett. \textbf{82}, 2430 (1999);
M.B. Smy (Super-Kamiokande Coll.),
hep-ex/9903034.

\bibitem{LSND}
C. Athanassopoulos \textit{et al.} (LSND Coll.),
Phys. Rev. Lett. \textbf{75}, 2650 (1995);
Phys. Rev. Lett. \textbf{77}, 3082 (1996);
Phys. Rev. Lett. \textbf{81}, 1774 (1998).

\bibitem{BP98}
J.N. Bahcall, S. Basu and M.H. Pinsonneault,
Phys. Lett. \textbf{B433}, 1 (1998), astro-ph/9805135.

\bibitem{BGG-98-review}
S.M. Bilenky, C. Giunti and W. Grimus,
hep-ph/9812360,
to be published in Progress in Particle and Nuclear Physics, Volume 43.

\bibitem{sun-analysis}
J.N. Bahcall, P.I. Krastev and A.Yu. Smirnov,
Phys. Rev. \textbf{D58}, 096016 (1998);
Y. Fukuda \textit{et al.},
Phys. Rev. Lett. \textbf{82}, 1810 (1999);
V. Barger and K. Whisnant,
Phys. Lett. \textbf{B456}, 54 (1999);
M.C. Gonzalez-Garcia \textit{et al.},
hep-ph/9906469.

\bibitem{LSND-check}
Booster Neutrino Experiment (BooNE), http://{\-}nu1.lampf.lanl.gov/{\-}BooNE;
I-216 $\nu_\mu\to\nu_e$ proposal at CERN,
http://{\-}chorus01.{\-}cern.ch/\~{}pzucchel/{\-}loi/;
Oak Ridge Large Neutrino Detector, http://{\-}www.{\-}phys.{\-}subr.{\-}edu/{\-}orland/;
NESS: Neutrinos at the European Spallation Source,
http://{\-}www.{\-}isis.{\-}rl.{\-}ac.{\-}uk/{\-}ess/{\-}neut\%5Fess.{\-}htm.

\bibitem{reviews}
S.M. Bilenky and B. Pontecorvo,
Phys. Rep. \textbf{41}, 225 (1978);
S.M. Bilenky and S.T. Petcov,
Rev. Mod. Phys. \textbf{59}, 671 (1987);
C.W. Kim and A. Pevsner,
\textit{Neutrinos in Physics and Astrophysics},
Contemporary Concepts in Physics, Vol. 8,
Harwood Academic Press, Chur, Switzerland, 1993.

\bibitem{PDG98}
C. Caso \textit{et al.} (Particle Data Group),
Eur. Phys. J. \textbf{C3}, 1 (1998).

\bibitem{MSW}
S.P. Mikheyev and A.Yu. Smirnov,
Yad. Fiz. \textbf{42}, 1441 (1985)
[Sov. J. Nucl. Phys. \textbf{42}, 913 (1985)];
Il Nuovo Cimento \textbf{C9}, 17 (1986);
L. Wolfenstein,
Phys. Rev. \textbf{D17}, 2369 (1978);
\textit{ibid.} \textbf{20}, 2634 (1979).

\bibitem{Bugey-95}
B. Achkar \textit{et al.} (Bugey Coll.),
Nucl. Phys. \textbf{B434}, 503 (1995).

\bibitem{four-models}
J.T. Peltoniemi, D. Tommasini and J.W.F. Valle,
Phys. Lett. \textbf{B298}, 383 (1993);
E.J. Chun \textit{et al.},
\emph{ibid} \textbf{357}, 608 (1995);
S.C. Gibbons \textit{et al.},
\emph{ibid} \textbf{430}, 296 (1998);
B. Brahmachari and R.N. Mohapatra,
\emph{ibid} \textbf{437}, 100 (1998);
S. Mohanty, D.P. Roy and U. Sarkar,
\emph{ibid} \textbf{445}, 185 (1998);
J.T. Peltoniemi and J.W.F. Valle,
Nucl. Phys. \textbf{B406}, 409 (1993);
Q.Y. Liu and A.Yu. Smirnov,
\emph{ibid} \textbf{524}, 505 (1998);
D.O. Caldwell and R.N. Mohapatra,
Phys. Rev. \textbf{D48}, 3259 (1993);
E. Ma and P. Roy,
\emph{ibid} \textbf{52}, R4780 (1995);
S. Goswami,
\emph{ibid} \textbf{55}, 2931 (1997);
A.Yu. Smirnov and M. Tanimoto,
\emph{ibid} \textbf{55}, 1665 (1997);
N. Gaur \textit{et al.},
\emph{ibid} \textbf{58}, 071301 (1998);
E.J. Chun, C.W. Kim and U.W. Lee,
\emph{ibid} \textbf{58}, 093003 (1998);
K. Benakli and A.Yu. Smirnov,
Phys. Rev. Lett. \textbf{79}, 4314 (1997);
Y. Chikira, N. Haba and Y. Mimura,
hep-ph/9808254;
C. Liu and J. Song,
hep-ph/9812381;
W. Grimus, R. Pfeiffer and T. Schwetz,
hep-ph/9905320.

\bibitem{four-phenomenology}
J.J. Gomez-Cadenas and M.C. Gonzalez-Garcia,
Z. Phys. \textbf{C71}, 443 (1996), hep-ph/9504246;
V. Barger, T.J. Weiler and K. Whisnant,
Phys. Lett. \textbf{B427}, 97 (1998), hep-ph/9712495;
V. Barger, Y.B. Dai, K. Whisnant and B.L. Young,
Phys. Rev. \textbf{D59}, 113010 (1999), hep-ph/9901388;
C. Giunti,
hep-ph/9906275;
hep-ph/9906456;
S.M. Bilenky \textit{et al.},
hep-ph/9907234.

\bibitem{BGKP-96}
S.M. Bilenky, C. Giunti, C.W. Kim and S.T. Petcov,
Phys. Rev. \textbf{D54}, 4432 (1996), hep-ph/9604364.

\bibitem{BGG-AB-96}
S.M. Bilenky, C. Giunti and W. Grimus,
Proc. of \textit{Neutrino '96}, Helsinki, June 1996,
edited by K. Enqvist \textit{et al.},
p.~174,
World Scientific, 1997,
hep-ph/9609343;
Eur. Phys. J. \textbf{C1}, 247 (1998), hep-ph/9607372.

\bibitem{Okada-Yasuda-97}
N. Okada and O. Yasuda,
Int. J. Mod. Phys. A \textbf{12}, 3669 (1997), hep-ph/9606411.

\bibitem{BGG-bounds}
S.M. Bilenky, C. Giunti and W. Grimus,
Phys. Rev. \textbf{D57}, 1920 (1998), hep-ph/9710209.

\bibitem{BGG-CP}
S.M. Bilenky, C. Giunti and W. Grimus,
Phys. Rev. \textbf{D58}, 033001 (1998), hep-ph/9712537.

\bibitem{BGGS-BBN-98}
S.M. Bilenky, C. Giunti, W. Grimus and T. Schwetz,
hep-ph/9804421
[to be published in Astropart. Phys.].

\bibitem{Barger-variations-98}
V. Barger, S. Pakvasa, T.J. Weiler and K. Whisnant,
Phys. Rev. \textbf{D58}, 093016 (1998), hep-ph/9806328.

\bibitem{BGGS-AB-99}
S.M. Bilenky, C. Giunti, W. Grimus and T. Schwetz,
hep-ph/9903454
[to be published in Phys. Rev. D].

\bibitem{CDHS-CCFR-84}
F. Dydak \textit{et al.} (CDHS Coll.),
Phys. Lett. \textbf{B134}, 281 (1984);
I.E. Stockdale \textit{et al.} (CCFR Coll.),
Phys. Rev. Lett. \textbf{52}, 1384 (1984).

\end{thebibliography}
\end{document}